# Solar System Processes Underlying Planetary Formation, Geodynamics, and the Georeactor

J. Marvin Herndon

**Transdyne Corporation**
San Diego, CA 92131 USA

mherndon@san.rr.com
http://UnderstandEarth.com

**Abstract:** Only three processes, operant during the formation of the Solar System, are responsible for the diversity of matter in the Solar System and are directly responsible for planetary internal-structures, including planetocentric nuclear fission reactors, and for dynamical processes, including and especially, geodynamics. These processes are: (*i*) Low-pressure, low-temperature condensation from solar matter in the remote reaches of the Solar System or in the interstellar medium; (*ii*) High-pressure, high-temperature condensation from solar matter associated with planetary-formation by raining out from the interiors of giant-gaseous protoplanets, and; (*iii*) Stripping of the primordial volatile components from the inner portion of the Solar System by super-intense solar wind associated with T-Tauri phase mass-ejections, presumably during the thermonuclear ignition of the Sun. As described herein, these processes lead logically, in a causally related manner, to a coherent vision of planetary formation with profound implications including, but not limited to, (*a*) Earth formation as a giant gaseous Jupiter-like planet with vast amounts of stored energy of protoplanetary compression in its rock-plus-alloy kernel; (*b*) Removal of approximately 300 Earth-masses of primordial volatile gases from the Earth, which began Earth's decompression process, making available the stored energy of protoplanetary compression for driving geodynamic processes, which I have described by the new whole-Earth decompression dynamics and which is responsible for emplacing heat at the mantle-crust-interface at the base of the crust through the process I have described, called mantle decompression thermal-tsunami; and, (*c*) Uranium accumulations at the planetary centers capable of self-sustained nuclear fission chain reactions.

## 1. Introduction

Early in 1939, Hahn and Strassmann (1939) reported their discovery of neutron-induced nuclear fission. Just months later, Flügge (1939) speculated on the possibility that self-sustaining chain reactions might have taken place under natural conditions in uranium ore deposits. Kuroda (1956) used Fermi's nuclear reactor theory (Fermi, 1947) to demonstrate the feasibility that, two billion years ago or before, thick seams of uranium ore might have become critical and functioned as thermal neutron reactors moderated by ground water. Sixteen years passed before French scientists discovered in 1972 the first of several fossil remains of natural nuclear reactors at Oklo, in the Republic of Gabon, Africa (Bodu et al., 1972). These had operated about 1.8 billion years ago as thermal neutron reactors, in much the same manner as predicted by Kuroda (Maurette, 1976), and had also operated to some extent as fast neutron breeder reactors (Fréjacques et al., 1975; Hagemann et al., 1975).

There is evidence that certain planets contain internal energy sources. In 1969 astronomers discovered that Jupiter radiates to space more energy than it receives. Verification followed, indicating that not only Jupiter, but Saturn and Neptune as well each radiate approximately twice as much energy as they receive from the Sun (Aumann et al., 1969; Conrath et al., 1991). For two decades, planetary scientists could find no viable explanation for the internal energy sources in these planets and declared that "by default" (Stevenson, 1978) or "by elimination" (Hubbard, 1990) the observed energy must come from planetary formation about $4.5 \times 10^9$ years ago. In 1992, using Fermi's nuclear reactor theory, I demonstrated the feasibility for planetocentric nuclear fission reactors as the internal energy sources for the giant outer planets (Herndon, 1992). Initially, I considered only hydrogen-moderated thermal neutron reactors, but soon demonstrated the feasibility for fast neutron reactors as well, which admitted the possibility of planetocentric nuclear reactors in non-hydrogenous planets (Herndon, 1993, 1994, 1996).

It is known that the Earth has an internal energy source at or near the center of the planet that powers the mechanism that generates and sustains the geomagnetic field. In 1993, using Fermi's nuclear reactor theory, I demonstrated the feasibility of a planetocentric nuclear fission reactor as the energy source for the geomagnetic field (Herndon, 1993). Initially, I could only postulate that the georeactor, as it is called, would operate as a fast neutron breeder reactor over the lifetime of the Earth. Subsequent state-of-the-art numerical simulations, made at Oak Ridge National Laboratory, verified that the georeactor could indeed function over the lifetime of the Earth as a fast neutron breeder reactor and, significantly, would produce helium in the same range of isotopic compositions observed in oceanic basalts (Herndon, 2003; Hollenbach and Herndon, 2001).

Raghavan (2002) demonstrated the feasibility of using geo-antineutrinos as a means for verifying the existence of the georeactor. Why is verification extremely important? As noted by Domogatski et al. (2004), "Herndon's idea about georeactor located at the center of the Earth, if validated, will open a new era in planetary physics."

The purpose of this paper is to disclose the nature of Solar System processes that underlie planetary formation, geodynamics, and the georeactor. The processes revealed lead logically, in causally-related ways, to planetary compositions, internal structures, and the basis for the georeactor. The processes disclosed also lead to a new vision of global dynamics, called *whole-Earth decompression dynamics* (Herndon, 2004c, 2005b, c), as well as to a new concept of heat transport within the Earth, called *mantle decompression thermal-tsunami*, which emplaces heat at the base of the crust. In a broader sense, the processes revealed lead to a fundamentally different view of planetary formation than considered over the past four decades and to a new understanding of the genesis of the matter that comprises the Solar System.

## 2. Nature and Origin of Planetary Matter

The constancy in isotopic compositions of most of the elements of the Earth, the Moon, and the meteorites indicates formation from primordial matter of common origin. Primordial elemental composition is yet manifest and determinable to a great extent in the photosphere of the Sun. The less volatile rock-forming elements, present in the outer regions of the Sun, occur in nearly the same relative proportions as in chondritic meteorites, the relative elemental abundances being related, not to chemical properties, but to nuclear properties.

Chondrites differ somewhat from one another in their respective proportions of major elements (Jarosewich, 1990; Wiik, 1969), in their states of oxidation (Herndon, 1996, Urey and Craig, 1953), mineral assemblages (Mason, 1962), and oxygen isotopic compositions (Clayton, 1993); accordingly, they are grouped into three distinct classes: *enstatite*, *carbonaceous* and *ordinary*. Virtually all approaches to whole-Earth composition are based upon the idea that the Earth is similar in composition to a chondrite meteorite. A major controversy within the Earth sciences began more than six decades ago with the choice of chondrite type as being representative of the Earth (Herndon, 2005a).

Only three major rock-forming elements, iron (Fe), magnesium (Mg) and silicon (Si), together with combined oxygen (O) and sulfur (S), comprise at least 95% of the mass of each chondrite and, by implication, each of the terrestrial planets. These five elements, because of their great relative abundances, act as a buffer assemblage. Minor and trace elements provide a great wealth of detail, but are slaves to that buffer system and are insufficiently abundant to alter conclusions derived from the major elements.

For decades, the abundances of major elements ($E_i$) in chondrites have been expressed in the literature as ratios, usually relative to silicon ($E_i/Si$) and occasionally relative to magnesium ($E_i/Mg$). By expressing Fe-Mg-Si elemental abundances as molar ratios relative to iron ($E_i/Fe$), as shown in Figure 1, I discovered a fundamental relationship bearing on the nature of chondrite matter that can be understood at different levels (Herndon, 2004b). In Figure 1, chondrite data points scatter about three distinct, well defined, least squares fit, straight lines, unique to their classes, despite mineralogical differences observed among members within a given class of chondrites.

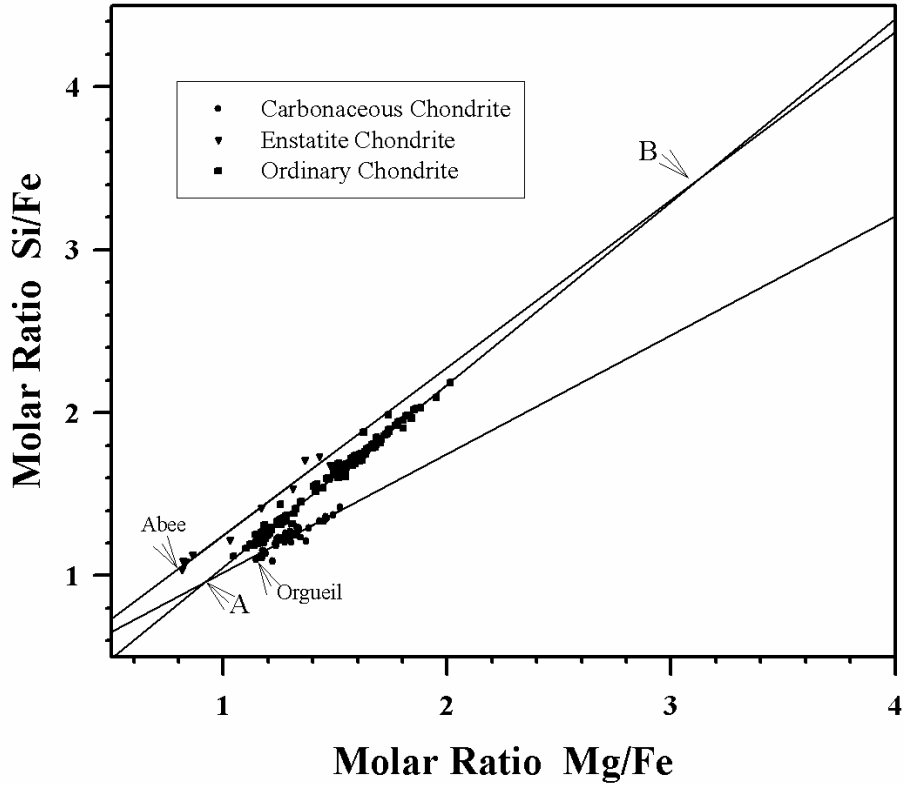

**Figure 1.** Molar (atom) ratios of Mg/Fe and Si/Fe from analytical data on 10 enstatite chondrites, 39 carbonaceous chondrites, and 157 ordinary chondrites. Data from (Baedecker and Wasson, 1975; Jarosewich, 1990; Wiik, 1969). Members of each chondrite class data set scatter about a unique, linear regression line. The locations of the volatile-rich Orgueil carbonaceous chondrite and the volatile-rich Abee enstatite chondrite are indicated. Line intersections A and B represent the compositions, respectively, of the primitive component and the partially-differentiated-enstatite-chondrite-like component from which the ordinary chondrites appear to have formed.

At one level of understanding, Figure 1 means that the well-mixed primordial matter became, or evolved to become, only three distinct types of matter which still retain more-or-less the full complement of readily condensable elements and which became the building blocks of the terrestrial planets. At a deeper level, as discussed in reference (Herndon, 2004b), the relationship shown in Figure 1 admits the possibility of ordinary chondrites having been derived from mixtures of two components, representative of the other two types of matter, mixtures of a relatively undifferentiated carbonaceous-chondrite-like *primitive* component and a partially differentiated enstatite-chondrite-like *planetary* component.

The interest here is not simply to understand the origin of chondrite meteorites, but to understand the nature of the physical processes leading to the evolution of their components from the well-mixed primordial progenitor material. The components of chondrite meteorites are in a sense like the results of experiments made in a laboratory, but absent knowledge of exact experimental conditions. Making sense out of these data can lead to a broader understanding of what processes are possible and impossible in the medium from which the planets formed.

The Abee enstatite chondrite and the Orgueil carbonaceous chondrite typify the primitive (least differentiated) end members of their respective types of matter, as shown in Figure 1. In terms of their elemental compositions, including their respective complements of volatile trace elements, they are virtually identical meteorites, an indication of a relatively simple chemical progression from their essentially uniform, well-mixed primordial parent matter. But these two meteorites are strikingly different in terms of their states of oxidation, mineral compositions, evidence of thermal exposure, and formation-location in the Solar System.

There have long been mainly two ideas about how the planets of the Solar System formed. In the 1940s and 1950s, the idea was discussed about planets "raining out" from inside of giant gaseous protoplanets with hydrogen gas pressures on the order of $10^2$-$10^3$ bar (Eucken, 1944; Kuiper, 1951a; Urey, 1951). But, in the early 1960s, scientists instead began thinking of primordial matter, not forming dense protoplanets, but rather spread out into a very low-density "solar nebula" with hydrogen gas pressures on the order of $10^{-5}$ bar. The idea of low-density planetary formation, often referred to as the standard model, envisioned that dust would condense at fairly low temperatures, and then would gather into progressively larger grains, and become rocks, then planetesimals, and ultimately planets (Stevenson, 1982; Wetherill, 1980).

These two ideas about planetary formation embody fundamentally different condensation processes which, I submit, are the underlying cause for the two unique types of chondritic matter shown in Figure 1. The immediate implication is that both processes were operant during the formation of the Solar System. The relative extent and region of each process can be ascertained to some certitude from thermodynamic considerations together with planetary data. Even within present limitations, a consistent picture emerges that is quite unlike the standard model of Solar System formation.

### 3. Low-Temperature, Low-Pressure Condensation

Following the publication by Cameron (1963) of his diffuse solar nebula models at pressures of about $10^{-5}$ bar, confusion developed during the late 1960s and early 1970s about the nature of the products anticipated to result by condensation from an atmosphere of solar composition at such low pressures. The so-called "equilibrium condensation" model was contrived and widely promulgated (Larimer and Anders, 1970). That model was predicated upon the later refuted assumption (Herndon 1978; Herndon and Suess, 1977) that the mineral assemblage characteristic of ordinary chondrite meteorites formed as the condensate from a gas of solar composition at pressures of about $10^{-5}$ bar.

The great majority of chondrites observed falling to Earth are called ordinary chondrites, the name denoting their great frequency of occurrence, ~98%. In terms of the five major elements comprising about 95% of the mass of each ordinary chondrite, their mineral assemblage is quite simple, as shown in Table 1. Silicon and magnesium occur combined with oxygen in the silicate minerals, olivine, $(MgO,FeO)_2SiO_2$, and pyroxene, $(MgO,FeO)SiO_2$. Some iron occurs combined with oxygen in the silicate minerals, some as iron metal, Fe, and some combined with sulfur as troilite, FeS. The minerals of ordinary chondrites are generally crystalline and typically show evidence of exposure to elevated temperatures.

**Table 1**. The mineral assemblages characteristic of chondritic meteorites. The hydrous C1 carbonaceous chondrites have a state of oxidation characteristic of low-pressure condensation to low temperatures. The highly-reduced enstatite chondrites are similar to the matter of the endo-Earth, the inner 82% of the Earth.

### *HYDROUS CHONDRITES*

| *Chondrite Type* | *Major Minerals* |
|---|---|
| **Carbonaceous Chondrites** | complex hydrous layer lattice silicate<br>  e.g. $(Mg, Fe)_6Si_4O_{10}(O, OH)_8$<br>epsomite, $MgSO_4 \cdot 7H_2O$<br>magnetite, $Fe_3O_4$ |

### *ANHYDROUS CHONDRITES*

| *Chondrite Type* | *Major Minerals* |
|---|---|
| **Carbonaceous Chondrites** | olivine, $(Fe, Mg)_2SiO_4$<br>pyroxene, $(Fe, Mg)SiO_3$<br>pentlandite, $(Fe, Ni)_9S_8$<br>troilite, FeS |
| **Ordinary Chondrites** | olivine, $(Fe, Mg)_2SiO_4$<br>pyroxene, $(Fe, Mg)SiO_3$<br>troilite, FeS<br>metal, (Fe-Ni alloy) |
| **Enstatite Chondrites** | pyroxene, $MgSiO_3$<br>complex mixed sulfides<br>  e.g. $(Ca, Mg, Mn, Fe)S$<br>metal, (Fe, Ni, Si alloy)<br>nickel silicide, $Ni_2Si$ |

Suess and I showed that the oxidized-iron content of ordinary-chondrite-silicate-minerals was consistent, not with their condensation from an atmosphere of solar composition, but from an atmosphere where hydrogen was about one-thousandth as abundant (Herndon and Suess, 1977). Subsequently, I showed (*i*) that there is at most only a single temperature, if any at all, where the ordinary chondrite mineral assemblage can exist in equilibrium with solar matter, and (*ii*) that condensation of that mineral assemblage would necessitate an atmosphere depleted in oxygen, as well as hydrogen, relative to solar matter (Herndon, 1978). The ordinary chondrite mineral assemblage is not the condensate from an atmosphere of solar composition at hydrogen pressures on the order of $10^{-5}$ bar. So, what then is the mineral assemblage expected?

From thermodynamic considerations it is possible to make some generalizations related to the condensation process in an atmosphere of solar composition. In that medium, the oxygen fugacity is dominated by the gas-phase reaction $H_2 + \frac{1}{2}O_2 = H_2O$ which is a function of temperature, but is essentially independent of pressure over a wide range of pressures where ideal gas behavior is approached. Oxygen fugacity controls the condensate state of oxidation at a particular temperature. At high temperatures the state of oxidation is extremely reducing, while at low temperatures it is quite oxidizing. The state of oxidation of the condensate ultimately becomes fixed at the temperature at which reaction with the gas phase ceases and/or equilibrium is frozen-in by the separation of gases from the condensate.

Condensation of an element or compound is expected to occur when its partial pressure in the gas becomes greater than its vapor pressure. Generally, at high pressures in solar matter, condensation is expected to commence at high temperatures. At low pressures, such as a hydrogen pressure of $10^{-5}$ bar, condensation is expected to progress at relatively low temperatures at a fairly oxidizing range of oxygen fugacity. At low temperatures, all of the major elements in the condensate may be expected to be oxidized because of the great abundance of oxygen in solar matter, relative to the other major condensable elements. Beyond these generalizations, in this low-pressure regime, precise theoretical predictions of specific condensate compounds may be limited by kinetic nucleation dynamics and by gas-grain temperature differences arising because of the different mechanisms by which gases and condensate lose heat.

Among the thousands of known chondrites, only a few, like the Orgueil carbonaceous chondrite, have a state of oxidation and mineral components with characteristics similar to those which might be expected as a condensate from solar matter at low pressures. Essentially all of the major elements in these few chondrites are oxidized, as shown in Table 1. The major silicate is not a well-defined crystalline phase like olivine, but is, instead, poorly-characterized phyllosilicate, a layer-lattice, claylike, hydrous material. The presence of sharp, angular shards of crystalline olivine and pyroxene in Orgueil (Reid et al., 1970) appear to be an admixed xenolithic component and shows no indication of alteration, suggesting the phyllosilicate is primary, rather than a secondary aqueous alteration product of olivine. Iron occurs, not as metal, but as magnetite, $Fe_3O_4$, (Hyman et al., 1978) which presents in a variety of unique morphologies including plaquettes and framboids (Hua and Buseck, 1998; Jedwab, 1971). In Orgueil sulfur occurs mainly as epsomite, $MgSO_4 \cdot 7H_2O$, (Endress and Bischoff, 1993) instead of as troilite, FeS.

There is debate as to how much alteration might or might not have occurred on the Orgueil meteorite's parent body (Tomeoka and Buseck, 1988). Nevertheless, that meteorite is the closest chondrite representative to what may be expected as a low-temperature, low-pressure condensate from the oxygen-rich gas of solar composition. Re-melting and/or re-evaporating and re-condensing Orgueil-like matter, after loss of primordial gases, may be expected to yield crystalline minerals, such as olivine and pyroxene, similar in composition to some other, more evolved, carbonaceous chondrites, such as the Allende meteorite which contains so much oxidized iron in its crystalline silicates, that there is very little remaining as the metal. Significantly, reflectance spectroscopy results appear to identify carbonaceous chondrite-like matter on the surfaces of bodies in the Kuiper Belt in the outer regions of the Solar System (Lederer and Vilas, 2003).

The idea of planetary formation from a diffuse solar nebula, with hydrogen pressures on the order of $10^{-5}$ bar, envisioned that dust would condense at fairly low temperatures, and then would gather into progressively larger grains, and become rocks, then planetesimals, and ultimately planets. In the main, that idea leads to the contradiction of the terrestrial planets having insufficiently massive cores, because the condensate would be far too oxidized for a high proportion of iron metal to exist. But as evidenced by Orgueil and similar meteorites, such low-temperature, low-pressure condensation did in fact occur, perhaps only in the evolution of matter of the outer regions of the Solar System, and thus may contribute to terrestrial planet formation only as a component of late addition veneer.

## 4. High-Temperature, High-Pressure Condensation

In 1944, on the basis of thermodynamic considerations, Eucken (1944) suggested core-formation in the Earth as a consequence of successive condensation from solar matter, on the basis of volatility, from the central region of a hot, gaseous protoplanet with molten iron metal first raining out at the center. Except for a few investigations initiated in the early 1950s (Bainbridge, 1962; Kuiper 1951a, 1951b; Urey, 1952), that idea languished when interest was diverted to Cameron's low-pressure solar nebula models (Cameron, 1963).

The enstatite chondrites consist of the most highly reduced natural mineral assemblage known (Table 1). The principal silicate mineral, enstatite, $MgSiO_3$, contains very little oxidized iron. The metal phase contains elemental silicon; magnesium and calcium, strongly lithophile (oxyphile) elements, occur in part as sulfides. And, unique nitrogen-containing minerals occur. The Abee enstatite chondrite has virtually the same relative abundance of volatile elements, such as lead and thallium, as the Orgueil carbonaceous chondrite, which consists of hydrous low-temperature minerals. But, in striking contrast, the Abee meteorite shows evidence of having been at melt or near-melt temperatures as evidenced by sub-euhedral crystals of enstatite embayed by iron metal. Interestingly, as Rudee and I have shown by metallurgical experiments, the Abee enstatite chondrite cooled from 700°C to 200°C in a matter of about 2 hours (Herndon and Rudee, 1978; Rudee and Herndon, 1981).

The formation of enstatite chondrites has posed something of an enigma for those who make models because, for low-temperature condensation at hydrogen pressures of about $10^{-5}$ bar, solar matter is much too oxidizing for that mineral assemblage. This has led to the suggestion that loss of $H_2O$ or $C/O \geq 0.9$ in solar matter might account for the state of reduction observed (Larimer, 1968).

On the basis of thermodynamic considerations, Suess and I showed at the high-temperatures for condensation at high-pressures, solar matter is sufficiently reducing, *i.e.*, it has a sufficiently low oxygen fugacity, for the stability of some enstatite chondrite minerals. However, formation of enstatite-chondrite-like condensate would necessitate thermodynamic equilibria being frozen-in at near-formation temperatures (Herndon and Suess, 1976). There is much to verify and learn about the process of condensation from near the triple point of solar matter, but the glimpses Suess and I have seen are remarkably similar to the vision of Eucken (1944), *i.e.*, molten iron raining out in the center of a hot, gaseous protoplanet.

At present, there is no adequate published theoretical treatment of solar-matter condensation from near the triple-point. But from thermodynamic and metallurgical considerations, some generalizations can be made. At the high temperatures at which condensation is possible at high pressures, nearly everything reacts with everything else and nearly everything dissolves in everything else. At such pressures, molten iron, together with the elements that dissolve in it, is the most refractory condensate.

There are reasons to associate the highly reduced matter of enstatite chondrites with the inner regions of the Solar System: (*i*) The regolith of Mercury appears from reflectance spectrophotometric investigations (Vilas, 1985) to be virtually devoid of FeO, like the silicates of the enstatite chondrites (and unlike the silicates of other types of chondrites); (*ii*) E-type asteroids (on the basis of reflectance spectra, polarization, and albedo), the presumed source of enstatite meteorites, are, radially from the Sun, the inner most of the asteroids (Zellner et al., 1977); (*iii*) Only the enstatite chondrites and related enstatite achondrites have oxygen isotopic compositions indistinguishable from those of the Earth and the Moon (Clayton, 1993); and, (*iv*) Fundamental mass ratios of major parts of the Earth (geophysically determined) are virtually identical to corresponding (mineralogically determined) parts of certain enstatite chondrites, especially the Abee enstatite chondrite (Herndon 1980, 1993, 1996).

In the absence of evidence to the contrary, the observed enstatite-chondritic composition of the terrestrial planets permits the deduction that these planets formed by raining out from the central regions of hot, gaseous protoplanets (Herndon, 2004d). With the possible exception of Mercury, the outer veneer of the terrestrial planets may contain other components derived from carbonaceous-chondrite-like matter and from ordinary-chondrite-like matter.

## 5. Evidence of Earth Being Like an Enstatite Chondrite

Imagine melting a chondrite in a gravitational field. At elevated temperatures, the iron metal and iron sulfide components will alloy together, forming a dense liquid that will settle beneath the silicates like steel on a steel-hearth. The Earth is like a spherical steel-hearth with a fluid iron-alloy core surrounded by a silicate mantle.

The Earth's core comprises about 32.5% by mass of the Earth as a whole. Only the enstatite chondrites, not the ordinary chondrites, have the sufficiently high proportion of iron-alloy that is observed for the core of the Earth, as shown in Figure 2. Moreover, other components of the interior of the Earth can be identified with corresponding components of an enstatite chondrite meteorite.

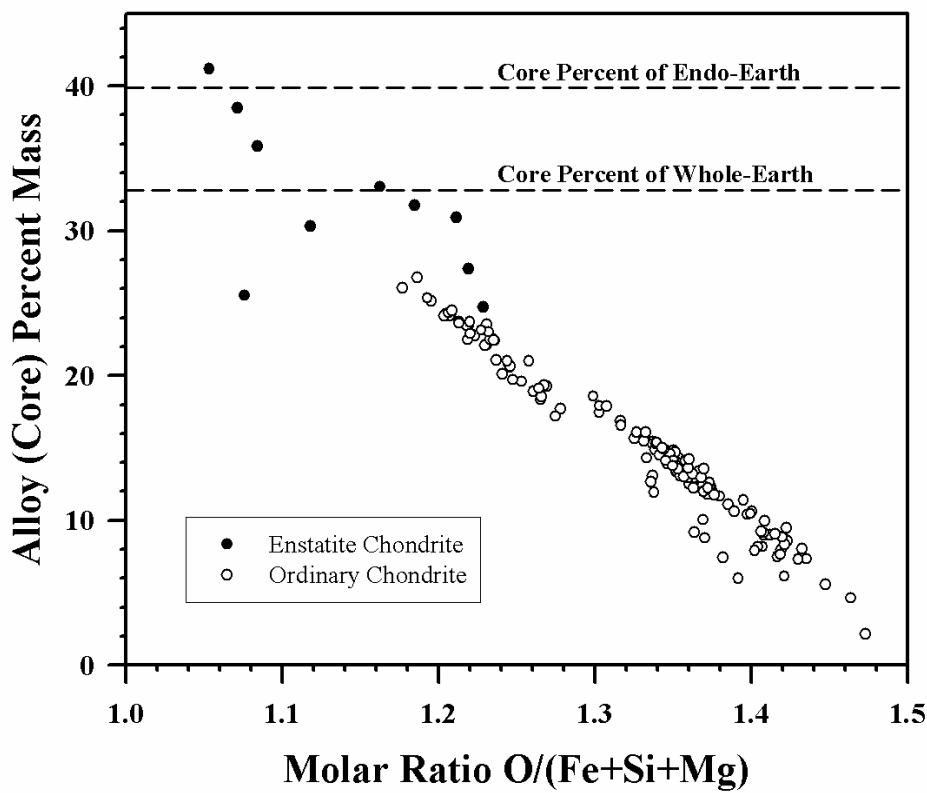

**Figure 2.** The percent mass of the alloy component of each of nine enstatite chondrites and 157 ordinary chondrites. This figure clearly shows that, if the Earth is chondritic in composition, the Earth as a whole, and especially the endo-Earth, is like an enstatite chondrite and *not* like an ordinary chondrite. The reason is clear from the abscissa which shows the molar ratio of oxygen to the three major elements with which it combines in enstatite chondrites and in ordinary chondrites. This figure also clearly shows that, if the Earth is chondritic in composition, the Earth as a whole, and especially the endo-Earth, has a state of oxidation like an enstatite chondrite and *not* like an ordinary chondrite. Data from (Baedecker and Wasson, 1975; Jarosewich, 1990; Kallemeyn et al., 1989; Kallemeyn and Wasson, 1981).

Oldham (1906) discovered the Earth's core by determining that beneath the crust the velocities of earthquake-waves increase with increasing depth, but only to a particular depth, below which their velocities abruptly and significantly become slower as they enter the core. When earthquake waves enter and leave the core, they change speed and direction. Consequently, there is a region at the surface, called the shadow zone, where earthquake-waves should be undetectable. But in the early 1930s, earthquake-waves were in fact detected in the shadow zone. Lehmann (1936) discovered the inner core by showing that a small solid object, within the fluid core, could cause earthquake waves to be reflected into the shadow zone.

Four years after its discovery by Inge Lehmann, Birch (1940) pronounced the composition of the inner core to be partially crystallized nickel-iron metal. Birch envisioned the Earth to be like an ordinary chondrite meteorite, the most common type of meteorite observed to fall to Earth. In arriving at that vision, Birch considered neither the rare, oxygen-rich carbonaceous chondrites, which contain little or no iron metal, nor the rare oxygen-poor enstatite chondrites, which contain iron metal and also some strange minerals, such as oldhamite, CaS, that are not found in the surface regions of the Earth.

Birch thought that nickel and iron were always alloyed in meteorites and he knew that the total mass of all elements heavier than nickel was too little to comprise a mass as large as the inner core. Birch therefore suggested that the inner core was nickel-iron metal that had begun to crystallize from the melt.

Nearly four decades later, I realized that elemental silicon, discovered in the 1960s in the metal of enstatite chondrites (Ringwood, 1961) under appropriate conditions could cause nickel to precipitate as nickel silicide, an intermetallic compound of nickel and silicon, like the mineral perryite, which had been discovered in the 1960s in enstatite chondrites (Ramdohr, 1964). The abstract of my 1979 paper (Herndon, 1979) states in its totality: "From observations of nature the suggestion is made that the inner core of the Earth consists not of nickel-iron metal but of nickel silicide".

After an inspiring conversation with Inge Lehmann in 1979, I progressed through the following logical exercise: If the inner core is in fact nickel silicide, then the Earth's core must be like the alloy portion of an enstatite chondrite. If the Earth's core is in fact like the alloy portion of an enstatite chondrite, then the Earth's core should be surrounded by a silicate shell like the silicate portion of an enstatite chondrite. This silicate shell, if it exists, should be bounded by a seismic discontinuity, because the silicates of enstatite chondrites have a different and more highly reduced composition than rocks that appear to come from within the Earth's upper mantle (Jagoutz et al., 1979). Using the alloy to silicate ratio of the Abee enstatite chondrite and the mass of the Earth's core, by simple ratio proportion I calculated the mass of that silicate shell. From tabulated mass distributions (Dziewonski and Gilbert, 1972), I then found that the radius of that predicted seismic boundary lies within about 1.2% of the radius at the seismic discontinuity that separates the lower mantle from the upper mantle. This logical exercise led me to discover the fundamental quantitative mass ratio relationships connecting the interior parts of the Earth with parts of the Abee enstatite chondrite that are shown in Table 2 (Herndon, 1980).

**Table 2**. Fundamental mass ratio comparison between the endo-Earth (core plus lower mantle) and the Abee enstatite chondrite (Herndon, 1980).

| Fundamental Earth Ratio | Earth Ratio Value | Abee Ratio Value |
|---|---|---|
| lower mantle mass to total core mass | 1.49 | 1.43 |
| inner core mass to total core mass | 0.052 | *theoretical* 0.052 if $Ni_3Si$ 0.057 if $Ni_2Si$ |
| inner core mass to (lower mantle+core) mass | 0.021 | 0.021 |

Discovery of the Mohorovičić discontinuity separating the crust from the mantle as well as discovery of the Earth's core and inner core in the first half of the 20th Century resulted from pronounced differences in seismic observables, whereas initially the mantle appeared to be uniform. In the 1960s, improvements in seismic resolution began to indicate difficult-to-observe discontinuities within the mantle (Stacey, 1969). These were initially assumed to result from pressure-induced crystal structure changes, rather than compositional boundaries.

From terrestrial seismic data (Dziewonski and Anderson, 1981; Dziewonski and Gilbert, 1972) the gross features of the inner 82% of the Earth, the lower mantle and core, collectively called the endo-Earth, appear to be relatively simple, consistent with the identification of that part being like an enstatite chondrite (Herndon, 1980, 1982). The upper mantle, on the other hand, displays several seismic discontinuities suggestive of different layers. The oxidized iron content (FeO) of primitive, ultramafic, upper-mantle-derived nodules (Jagoutz et al., 1979) would be out of equilibrium if in contact with the virtually FeO-free $MgSiO_3$ lower mantle, implying one or more layers of yet unknown but chemically different composition within the upper mantle. Such layering is consistent with the addition of carbonaceous-chondrite-like matter and/or ordinary-chondrite-matter during the latter stages of Earth formation (Wetherill, 1980). Indeed, just such a chondritic component is discernable in primitive ultramafic, upper-mantle-derived nodules (Jagoutz et al., 1979).

## 6. Overview of Solar System Formation

To understand more clearly the implications arising from protoplanetary Earth formation, it is helpful to envision the overall environment as indicated by chondrite chemical evidence and observational data. Although there is an evolutionary pre-history to the origin of the Solar System, involving among other things element nucleosynthesis, that pre-history is not considered here.

There seems to be little doubt that the oxidized, hydrous carbonaceous chondrites, like Orgueil, originate in the outer reaches of our Solar System, regions sufficiently cold to permit the retention of water in the vacuum of space for billions of years. The oxidation state of Orgueil-like carbonaceous chondrites is just what one would expect for solar-matter low pressure condensation at low temperatures.

The highly-reduced matter of the inner regions of the Solar System, on the other hand, appears to have originated quite differently. In the main, the terrestrial planets are like the highly-reduced enstatite chondrite meteorites. Thermodynamic considerations are consistent with the concept of Eucken (1944) that the terrestrial planets, like the Earth, rained out from the central regions of hot, gaseous protoplanets.

From solar abundances (Anders and Grevesse, 1989), the mass of protoplanetary-Earth was 275-305$m_E$, not very different from the mass of Jupiter, 318$m_E$. The formation of early-phase close-in gas giants in our own planetary system, is certainly consistent with observations and implications of near-to-star giant gaseous planets in other planetary systems (Fischer and Valenti, 2005; Santos et al., 2003; Udry et al., 2003), so it is no longer necessary to assume planet migration to explain those observations.

Solar primordial gases and volatile elements were separated from the terrestrial planets soon after planet formation, presumably early during some solar super-luminous event, such as the T-Tauri phase mass-ejections, presumably associated with the thermonuclear ignition of the Sun (Herbig, 1962; Joy, 1945; Lada, 1985; Lehmann et al., 1995). Indeed, there is some reason to think that Mercury was only partially formed at the time of super-luminosity.

As I discussed (Herndon, 2004b), the relationship shown in Figure 1 admits the possibility of ordinary chondrites having been derived from mixtures of two components, representative of the other two types of matter, mixtures of a relatively undifferentiated carbonaceous-chondrite-like primitive component and a partially differentiated enstatite-chondrite-like component. All ordinary chondrites are depleted relative to solar matter in siderophile refractory elements, such as iridium and osmium. Siderophile refractory element depletion in individual ordinary chondrites, as I have shown, is proportional to their relative respective proportion of the partially differentiated enstatite-chondrite-like component, indicating a single reservoir source of partially differentiated enstatite chondrite-like matter (Herndon, 2004b).

The high bulk density of planet Mercury indicates that much of the silicate matter for the upper portion of Mercury's mantle was lost at some previous time (Bullen, 1952; Urey, 1951, 1952). I have suggested that some matter from the protoplanet of Mercury, Mercury's complement of lost elements, became that partially differentiated enstatite-chondrite-like planetary component of the ordinary chondrites, presumably separated during the time of Mercury's core formation through dynamic instability and/or expulsion during the Sun's initially violent ignition and approach toward thermonuclear equilibrium. I have suggested that the Mercurian component was then re-evaporated together with a more oxidized component of primitive matter and ended up mainly in the asteroid belt, the presumed source-region for the ordinary chondrites (Chapman, 1996). Such a picture would seem to explain for the ordinary chondrites, their major element compositions, their intermediate states of oxidation, and their ubiquitous deficiencies of refractory siderophile elements, and would explain as well a major, primary source of matter in the asteroid belt.

The approximately seven-fold greater depletion of refractory siderophile elements, within the ordinary chondrites' partially differentiated enstatite chondrite-like planetary component, than other, more volatile, siderophile elements such as nickel, cobalt, and gold, indicates that planetary-scale differentiation, at least in this one instance, progressed in a heterogeneous manner (Herndon, 2004a, b, e).

Although the terrestrial planets appear to have rained out from the central regions of hot, gaseous protoplanets, evidence suggests some outer, minor, secondary accretion of oxidized matter in the grain-growth accumulation way envisioned by, for example, Wetherill (1980). Such secondary accumulation may consist in the main of carbonaceous chondrite-like matter, ordinary chondrite-like matter, and their derivatives, for example, iron meteorites and achondrites. I have estimated that the total mass of ordinary chondrite matter originally present in the Solar System amounts to only $1.83 \times 10^{24}$ kg (Herndon, 2004e). That amount of mass is insufficient to form a planet as massive as the Earth, but may have contributed significantly to the formation of Mars, as well as adding to the veneer of other planets, including the Earth. Presently, only about 0.1% of that mass remains in the asteroid belt.

### 7. Implications of Protoplanetary Earth Formation

The principal consequences of Earth's origin from within a giant gaseous protoplanet are profound and affect virtually all areas of geophysics in major, fundamental ways. Principal implications result (*i*) from Earth having been compressed by about 300 Earth-masses of primordial gases, and (*ii*) from the deep-interior having a highly-reduced state of oxidation. The former provides Earth's main geodynamic driving-energy and leads to a new vision of global dynamics, which I call whole-Earth decompression dynamics (Herndon, 2005b, c) and which, among other things, leads to new geophysical concepts related to heat emplacement at the base of the crust. The latter results in great quantities of uranium and thorium existing within the Earth's core, and leads to the feasibility of the georeactor, a hypothesized natural, nuclear fission reactor at the center of the Earth as the energy source for the geomagnetic field.

## 8. Evidence of Earth as a Jupiter-Like-Gas-Giant

Planets generally consist of more-or-less uniform, closed, concentric shells of matter, layered according to density. The crust of the Earth, however, is an exception. Approximately 29% of the surface area of the Earth is composed of the portions of continents that presently lie above mean sea level; an additional 12% of the surface area of the Earth is composed of the continental margins, which are submerged to depths of no more than 2 km (Mc Lennan, 1991). The continental crust is less dense and different in composition than the remaining surface area, which is composed of ocean-floor basalt.

To date there has been no adequate geophysical explanation to account for the formation of the non-contiguous, crustal continental rock layer, except the idea put forth by Hilgenberg (1933) that in the distant past for an unknown reason the Earth had a smaller diameter and, consequently, had a smaller surface area. From modern surface area measurements, I calculated that the smaller radius required would be about 64% of its current radius, which would yield a mean density for the Earth of 21 g/cm$^3$. The reason for Earth's smaller radius, I submit, is that the Earth rained out from within a giant gaseous protoplanet and originally formed as the rock-plus-alloy kernel of a giant gaseous planet like Jupiter (Herndon, 2004c, 2005c).

The mass of protoplanetary-Earth, calculated from solar abundance data (Anders and Grevesse, 1989) by adding to the condensable-planetary elements their proportionate amount of solar elements that are typically gases (*e.g.*, H, He) or that form volatile compounds (*e.g.*, O, C, N), lies in the range of about 275 to 305 times the mass of the present-day Earth. That mass is quite similar to Jupiter's mass, 318$m_E$.

**Table 3.** Published model pressure and density estimates (Podolak and Cameron, 1974; Stevenson and Salpeter, 1976) at the gas-rock boundary of Jupiter, shown for comparison with theoretical calculation of compressed Earth density at the same pressures.

| Jupiter Model Pressure (Mbar) | Jupiter Model Density (g/cm$^3$) | Compressed Earth Density (g/cm$^3$) |
|:---:|:---:|:---:|
| 43 | 18 | 20 |
| 46 | 18 | 21 |
| 60 | 20 | 23 |

Pressures at the gas-rock boundary within the interior of Jupiter are estimated to be in the range from 43 Mbar to 60 Mbar (Podolak and Cameron, 1974; Stevenson and Salpeter, 1976). Using a theoretical Thomas-Fermi-Dirac approach (Salpeter and Zapolsky, 1967), I calculated density at Jupiter-model, gas-rock-boundary pressures for matter having the approximate composition of the Earth as a whole. The calculations are based upon eight chemical elements that account about 98% of the Earth's mass, assume volume additivity, and ignore phase separations and transitions. The results of the calculations, presented in Table 3, show that a Jovian-like gas envelope is sufficient to compress the protoplanetary alloy-plus-rock core that became the Earth to a mean density of 21 g/cm$^3$.

The density value of 21 g/cm$^3$, estimated to result from compression by the great mass of giant-planet gases, is identical to that expected for a smaller Earth with a contiguous, closed, crustal continental shell. That identity, I submit, stands as evidence of the Earth having been a giant, gaseous planet like Jupiter (Herndon, 2004c, d).

## 9. Whole-Earth Decompression Dynamics

Early in the 20$^{th}$ Century, Wegener (1912) proposed that the continents at one time had been united, but subsequently had separated and drifted through the ocean floor to their present positions. After being ignored for half a century, Wegener's idea of continental drift re-emerged, cast into a new form called plate tectonics theory, with more detail and with new supporting observational data.

In plate tectonics, oceanic basalt, observed erupting from the mid-oceanic ridges, is thought to creep slowly across the ocean basin and to subduct, to plunge into the Earth, typically into submarine trenches. This theory appears to explain many geologic features observed at the Earth's surface, such as magnetic striations on the ocean floor, but necessitates solid-state mantle convection (Davies, 1977; Peltier, 1989; Runcorn, 1965), for which there is no unambiguous evidence despite decades of investigations.

Hilgenberg (1933) published a fundamentally different idea about the continents. He imagined that the Earth, for an unknown reason, was initially smaller in diameter, without oceans, and that the continents formed a uniform shell of matter covering the entire surface of the planet. Hilgenberg's idea, that the Earth subsequently expanded, fragmenting the uniform shell of matter into the continents and creating ocean basins in between, is the basis for Earth expansion theory (Carey, 1976, 1988; Scalera, 1990; Scalera and Jacob, 2003).

The principal impediments to the idea of Earth expansion have been (*i*) the lack of knowledge of a mechanism that could provide the necessary energy (Beck, 1969; Cook and Eardley, 1961) without departing from the known physical laws of nature (Jordan, 1971) and (*ii*) the ocean floors are less than 200 million years old which would seem to imply very recent expansion. In 1982, Scheidegger stated concisely the prevailing view, "Thus, if expansion on the postulated scale occurred at all, a completely unknown energy source must be found" (Scheidegger, 1982). Recently, I disclosed just such an energy source that follows from fundamental considerations (Herndon, 2004c, 2005b, c), the energy of protoplanetary compression, and set forth a different geodynamic theory, called whole-Earth decompression dynamics, which unifies seemingly disparate elements of plate tectonics theory and Earth expansion theory into one self-consistent description of the dynamics of the Earth as a whole.

After being stripped of its great, Jupiter-like overburden of volatile protoplanetary constituents, presumably by the high temperatures and/or by the violent activity, such as T - Tauri phase solar wind (Herbig, 1962; Joy, 1945; Lada, 1985; Lehmann et al., 1995), associated with the thermonuclear ignition of the Sun, the Earth would inevitably begin to decompress, to rebound toward a new hydrostatic equilibrium. The initial whole-Earth decompression is expected to result in a global system of major *primary* cracks appearing in the rigid crust which persist and are identified as the global, mid-oceanic ridge system, just as explained by Earth expansion theory. But here the similarity with that theory ends. Whole-Earth decompression dynamics sets forth a different mechanism for whole-Earth dynamics which involves the formation of *secondary* decompression cracks and the in-filling of those cracks, a process which is not limited to the last 200 million years.

As the Earth subsequently decompresses and swells from within, the deep interior shells may be expected to adjust to changes in radius and curvature by plastic deformation. As the Earth decompresses, the area of the Earth's rigid surface increases by the formation of secondary decompression cracks often located near the continental margins and presently identified as submarine trenches. These secondary decompression cracks are subsequently in-filled with basalt, extruded from the mid-oceanic ridges, which traverses the ocean floor by gravitational creep, ultimately plunging into secondary decompression cracks, thus emulating subduction.

As viewed today from the Earth's surface, the consequences of whole-Earth decompression dynamics appear very similar to those of plate tectonics, but with some profound differences. In fact, most of the evidence usually presented in support of plate tectonics also supports whole-Earth decompression dynamics. Just as in plate tectonics, one sees seafloor being produced at the mid-oceanic ridge, slowly moving across the ocean basin and disappearing into the Earth. But unlike plate tectonics, the basalt rock is not being re-cycled continuously by convection; instead, it is simply in-filling secondary decompression cracks. From the surface it may be very difficult indeed to discriminate between plate tectonics and whole-Earth decompression dynamics.

Usually arrayed as supporting plate tectonics theory, observations of ocean-floor magnetic striations, transform faults, island arc formation, and the generation and distribution of earthquakes are, I submit, consequences of whole-Earth decompression dynamics. These have the same basis and understanding in whole-Earth decompression dynamics as in plate tectonics.

Moreover, mantle seismic tomography results can be interpreted as imaging in-filled decompression cracks (Bunge et al., 2003). Seismic differences that are used to arrive at such images are not necessarily a reflection of temperatures, as often assumed, but can arise from differences in densities and/or differences in compositions. Moreover, the images are static; motion is only inferred on the basis of anticipations.

But there are global, fundamental differences between whole-Earth decompression dynamics and plate tectonics, especially as pertains to the growth of ocean-floor, to the origin of oceanic trenches, to the fate of down-plunging slabs, to the displacement of continents, and to the emplacement of heat at the base of the crust.

### 10. Mantle Decompression Thermal-Tsunami

Previously in geophysics, only three heat transport processes have been considered: conduction, radiation, and convection or, more generally, buoyancy-driven mass transport. As a consequence of whole-Earth decompression dynamics, I add a fourth, called mantle decompression thermal-tsunami (Herndon, 2006).

As the Earth decompresses, heat must be supplied to replace the lost heat of protoplanetary compression. Otherwise, decompression would lower the temperature, which would impede the decompression process.

Heat generated within the core from actinide decay or fission or from radioactive decay within the mantle may enhance mantle decompression by replacing the lost heat of protoplanetary compression. The resulting decompression, beginning as low as at the bottom of the mantle, will tend to propagate throughout the mantle, like a tsunami, until it reaches the impediment posed by the base of the crust. There, crustal rigidity opposes continued decompression, pressure builds and compresses matter at the mantle-crust-interface, resulting in compression heating. Ultimately, pressure is released at the surface through volcanism and through secondary decompression crack formation and/or enlargement.

It has been long known through experience in deep mines and with bore-holes that temperature increases with depth within the crust. For more than half a century geophysicists have made measurements of continental and oceanic heat flow with the aim of determining the Earth's heat loss (Table 4). Pollack et al. (1993) estimate a global heat loss of 44.2 TW (1 TW=$10^{12}$ W) based upon 24,774 observations at 20,201 sites.

Previously, numerous attempts have been made to match measured global heat loss with radionuclide heat production from various geophysical models involved with plate tectonics. Usually, models are made to yield the very result they model, but in this case there is a problem. Current models rely upon radiogenic heat for geodynamic processes, geomagnetic field generation, and for the Earth's heat loss. The problem is that radionuclides cannot even satisfy just the global heat loss requirements.

Previous estimates of global heat production invariably come from the more-or-less general assumption that the Earth's current heat loss consists of the steady heat production from long-lived radionuclides ($^{235}$U, $^{238}$U, and $^{40}$K). Estimates of present-day global radiogenic heat production, based upon chondritic abundances, typically range from 19 TW to 31 TW. These represent an upper limit through the tacit assumption of rapid heat transport irrespective of assumed radionuclide locations. The short-fall in heat production, relative to Earth's measured heat loss (Pollack et al., 1993), has led to speculation that the difference might be accounted for by residual heat from Earth's formation, ancient radiogenic heat from a time of greater heat production, or, perhaps, from a yet unidentified heat source (Kellogg et al., 1999).

**Table 4**. Continental and oceanic mean heat flow and global heat loss. From Pollack et al. (1993)

| Reference | Continental Heat Flow $m$Wm$^{-2}$ | Oceanic Heat Flow $m$Wm$^{-2}$ | Global Heat Flow $m$Wm$^{-2}$ | Global Heat Loss $10^{12}$W |
|---|---|---|---|---|
| Williams et al. (1974) | 61 | 93 | 84 | 42.7 |
| Davies (1980) | 55 | 95 | 80 | 41.0 |
| Sclater et al. (1980) | 57 | 99 | 82 | 42.0 |
| Pollack et al. (1993) | 65 | 101 | 87 | 44.2 |

One of the consequences of Earth formation as a giant, gaseous, Jupiter-like planet (Herndon, 2004d), as described by whole-Earth decompression dynamics (Herndon, 2004c, 2005b, 2005c), is the existence of a vast reservoir of energy, the stored energy of protoplanetary compression, available for driving geodynamic processes related to whole-Earth decompression. Some of that energy, I submit, is emplaced as heat at the mantle-crust-interface at the base of the crust through the process of mantle decompression thermal-tsunami. Moreover, some radionuclide heat may not necessarily contribute directly to crustal heating, but rather to replacing the lost heat of protoplanetary compression, which helps to facilitate mantle decompression.

### 11. Precipitation of the Structures of the Endo-Earth

One of the consequences of Earth formation by raining out from the central regions of a hot, gaseous protoplanet is the highly reduced state of oxidation of its interior (Eucken, 1944; Herndon and Suess, 1976). The Earth consists in the main of two distinct reservoirs of matter separated by the seismic discontinuity that occurs at a depth of about 680 km and which separates the mantle into upper and lower parts (Herndon, 1980). The endo-Earth, the inner 82% of the Earth's mass consists of the highly reduced lower mantle and core; the more oxidized exo-Earth is comprised of the components of the upper mantle and crust.

The matter comprising the endo-Earth precipitated from primordial gases under conditions that severely limited its oxygen content, relative to its other elements (Eucken, 1944; Herndon, 2004d; Herndon and Suess, 1976). The oxidation state of the condensate determines not only the relative mass of the core, but the elements the core contains, and the compounds which precipitate from the core and that give it its structure and its energy production capability. The oxidation state of the core cannot be subsequently changed, even by the pressures that prevail in that region.

The seismically-deduced structure, divisions, and components of the endo-Earth are essentially identical to corresponding parts of the Abee enstatite chondrite meteorite, as shown by the mass ratio relationships presented in Table 2. The identity of the components of the Abee enstatite chondrite with corresponding components of the Earth (Herndon, 1980, 1993, 1998) means that with reasonable confidence one can understand the composition of the Earth's core by understanding the components of Abee meteorite or of one like it.

Envision highly reduced condensate, like that of the Abee enstatite chondrite and the endo-Earth, raining out from near the triple point of solar matter in the center of a hot giant gaseous protoplanet (Eucken, 1944; Herndon, 2004d; Herndon and Suess, 1976). The magnesium, silicon, oxygen, and sulfur of enstatite-chondritic-like protoplanetary matter may have all begun their condensate origin dissolved in iron metal, along with minor and trace elements. Because of the extremely low oxygen fugacity in that medium at the high temperatures at which condensation is possible at high pressures, the amount of oxygen in the multi-element condensate would have been severely limited, even though oxygen is more abundant than the sum of all of the readily condensable elements of solar matter.

After raining out in the center of a hot gaseous protoplanet, elements of the condensate would be expected to compete on the basis of chemical activity and, during cooling, would begin to precipitate from the liquid condensate forming the interior parts of the planet. The dominant factors governing subsequent precipitation are oxyphilicity (affinity for oxygen) and incompatibility.

Elements have different chemical affinities for oxygen, which are related to their different oxidation potentials. Generally, oxyphile elements of the initial multi-element protoplanetary condensate will compete for available oxygen and will separate from the iron-alloy like slag separates from steel on a steel-hearth.

In ordinary-chondrite matter, there is more than enough oxygen available for all oxyphile elements (including uranium and thorium) with some left over to combine with iron. Consequently, if the Earth as a whole really were like an ordinary chondrite meteorite, there would be no uranium and thorium in the core and the core would be too small (Figure 2). But that is not the case.

Highly reduced matter, like that of the Abee enstatite chondrite and the endo-Earth, was separated from primordial solar gases under conditions that severely limited the oxygen content (Eucken, 1944; Herndon, 2004d; Herndon and Suess, 1976). For the protoplanetary Earth, elements of the condensate with a high affinity for oxygen (oxyphile elements) would be expected to combine with the limited available oxygen to form, atop the iron-alloy core, a low-density silicate mantle of $MgSiO_3$ which, at lower mantle pressures, is stable in a perovskite crystal structure (Chaplot and Choudhury, 2001; Chaplot et al., 1998; Ito and Matsui, 1978).

As a consequence of its highly-reduced state of oxidation, the protoplanetary condensate that became the endo-Earth had insufficient oxygen to accommodate all of its oxyphile elements. As a consequence, certain oxyphile elements, including Si, Mg, Ca, U, and Th, occur in part in the iron-based alloy portion of the Abee enstatite chondrite and in the Earth's core. Oxyphile elements are generally incompatible in an iron-alloy and upon cooling these ultimately tend to precipitate as non-oxides, mainly as sulfides, at the earliest thermodynamically-feasible opportunity.

Based upon well-known metallurgical principles (Inoue and Suito, 1994; Ribound and Olette, 1978), the portion of calcium and magnesium, occurring in the core and being incompatible in an iron-based alloy, would be expected to combine with sulfur to form oldhamite, CaS, and niningerite, MgS, low-density, high-temperature precipitates, which would float to the outer surface of the fluid core. These CaS and MgS precipitates, as I have suggested (Herndon, 1993, 1996, 2005a), are responsible for the observed seismic "roughness" at the core-mantle boundary, called D''.

Upon further cooling, it is expected that dissolved silicon (Si) in the fluid core will combine with nickel (Ni) and precipitate as nickel silicide, which will settle by gravity, forming the Earth's solid inner core (Herndon, 1979, 1980, 2005a). As shown in Table 2, a fully crystallized nickel silicide inner core would have precisely the mass observed, thus providing strong supporting evidence.

## 12. Radionuclides of the Endo-Earth

For decades there has been much discussion as to the possible existence of $^{40}K$ in the Earth's core. Although there are some indications from enstatite meteorites of alloy-originated potassium, specifically in the mineral djerfisherite, $K_6(Cu, Fe, Ni)_{25}S_{26}Cl$, the relative proportion of non-oxide potassium appears to represent at most only a few percent of the potassium complement (Fuchs, 1966). In the Abee enstatite chondrite, most of the potassium occurs in the mineral plagioclase, $(Na, Ca)(Si, Al)_4O_8$, which would seem to suggest that most of the endo-Earth's $^{40}K$ occurs in the lower mantle, perhaps in the region near the boundary of the upper mantle. Additional investigations are needed to be any more precise regarding the distribution of $^{40}K$.

Although there may be some intrinsic uncertainty as to amount of $^{40}K$, if any, in the Earth's core, current data on the uranium distribution in enstatite chondrites clearly indicate the non-lithophile behavior of that element in EH/E4 enstatite chondrites, like the Abee meteorite, and, by inference, in the endo-Earth. Generally, uranium occurs within the mineral oldhamite, CaS, an indication that in the enstatite chondrite matter, uranium is a high-temperature precipitate. Chemical leaching experiments show that Abee-uranium behaves as a sulfide (Matsuda et al., 1972). The tentative assignment of uranium as the mono-sulfide, US, seems reasonable. As currently-available instrumental capability for determining this information quite precisely exists, I have recommended the requisite investigations (Herndon, 1998).

Within the Earth's core, one would expect uranium to precipitate at a high temperature. Just as uranium, a trace element, was swept-up or co-precipitated with a more abundant high-temperature precipitate, oldhamite, CaS, in enstatite chondrites, one might expect to some extent the possibility of a similar fate within the Earth's core. Ultimately, uranium, being the densest substance, would be expected to collect at the Earth's center. Unlike other trace elements such as thorium, uranium masses of at least ~1 kg occurring as nodules early in Earth's history would have been able to maintain sustained nuclear fission chain reactions that could generate sufficient heat to melt their way out of any mineral-occlusion impediment on their descent to the center of the Earth.

Russian scientists (Anisichkin et al., 2003; Rusov et al., 2004) have suggested the possibility of precipitated uranium accumulating in a layer atop the inner core and participating in a slow-burning nuclear-fission wave front reaction. To me, it seems that a uniform layer would be too thin, allowing too great a proportion of neutrons to escape for maintenance of criticality. But uniformity is only one possibility. In this remote and strange frontier, it is a good idea to keep an open mind on all of the possible georeactor variations.

Thorium, like uranium, occurs exclusively in the alloy portion of the Abee enstatite chondrite and by implication in the Earth's core. Also, thorium, like uranium, occurs in that meteorite within the mineral oldhamite, CaS (Murrell and Burnett, 1982), an indication of its being a high-temperature precipitate. Chemical leaching experiments indicate that Abee-thorium behaves in part as a sulfide, and in part as an unknown non-sulfide (Matsuda et al., 1972). Unlike uranium, accumulations of thorium would not have been able to sustain nuclear fission chain reactions.

Thus, it would appear that uranium and thorium may occur at the core-mantle boundary occluded in the core floaters, the low-density, high-temperature precipitate, oldhamite, CaS, atop the fluid core or, alternatively, they may be concentrated at the center of the Earth, depending upon respective precipitation and accumulation dynamics. Presently, there is no methodology by which to predict the relative proportion of these at the two boundaries of the core, its center and its surface. Because of the ability of ~1 kg nodules of uranium to undergo self-sustaining nuclear fission chain reactions, which can melt free of occlusion, one might expect uranium to occur primarily at the center of the Earth and thorium to occur at the core-mantle boundary within oldhamite.

## 13. Radionuclides of the Exo-Earth

It would be desirable to be able to specify the radionuclide distribution within the exo-Earth, the upper mantle and crust. But at present there is uncertainty in the compositions of the layers of the upper mantle and uncertainty as to the composition of the parent materials for that region of the Earth. Moreover, because of mantle decompression thermal-tsunami, measured heat loss from the crust can no longer be considered a justification for high-radionuclide content of the exo-Earth. As a "ball park" estimate, one might guess that the radionuclide complement of the exo-Earth represents an additional 18% of the endo-Earth complement, with much of the exo-Earth uranium and thorium residing in the crust. Ultimately, it should be possible to refine these estimates by tedious efforts to discover fundamental quantitative relationships that lead logically to that information.

## 14. Georeactor Nuclear Fission

Nuclear fission produces energy, consumes uranium, and produces neutron-rich fission products which subsequently β⁻ decay, yielding antineutrinos. Detection of georeactor-produced antineutrinos is one way to validate the existence of the georeactor (de Meijer et al., 2004; Domogatski et al., 2004; Fiorentini et al., 2004; Raghavan, 2002).

Using Fermi's nuclear reactor theory (Fermi, 1947), in 1993, I demonstrated the feasibility of a planetocentric nuclear fission reactor as the energy source for the geomagnetic field (Herndon, 1993). Initially, I could only postulate that the georeactor would operate as a fast neutron breeder reactor over the lifetime of the Earth (Herndon 1994, 1996). Subsequent state-of-the-art numerical simulations, made at Oak Ridge National Laboratory, verified that the georeactor could indeed function over the lifetime of the Earth as a fast neutron breeder reactor and, significantly, would produce helium in the same range of isotopic compositions observed in oceanic basalts (Herndon, 2003; Hollenbach and Herndon, 2001).

Georeactor numerical simulation calculations are made using the SAS2 analysis sequence contained in the SCALE Code Package from Oak Ridge National Laboratory (SCALE 1995) that has been developed over a period of three decades and has been extensively validated against isotopic analyses of commercial reactor fuels (England et al., 1984; Hermann and DeHart, 1998). The SAS2 sequence invokes the ORIGEN-S isotopic generation and depletion code to calculate concentrations of actinides, fission products, and activation products simultaneously generated through fission, neutron absorption, and radioactive decay. The SAS2 sequence performs the 1-D transport analyses at selected time intervals, calculating an energy flux spectrum, updating the time-dependent weighted cross-sections for the depletion analysis, and calculating the neutron multiplication of the system.

From nuclear reactor theory (Fermi, 1947), the defining condition for self-sustaining nuclear fission chain reactions is that $k_{eff}$ = 1.0. The value of $k_{eff}$ represents the number of fission neutrons in the current population divided by the number of fission neutrons in the previous population. If $k_{eff}$ > 1.0, the neutron population and the energy output are increasing and will continue until changes in the fuel, moderators, and neutron absorbers cause $k_{eff}$ to decrease to 1.0. If $k_{eff}$ < 1.0, the neutron population and energy output are decreasing and will eventually decrease to zero. If $k_{eff}$ = 1.0, the neutron population and energy output are constant.

Natural uranium consists mainly of the readily-fissionable $^{235}$U and the essentially non-fissionable $^{238}$U. In a natural reactor, the value of $k_{eff}$ is strongly dependent upon the ratio $^{235}$U/$^{238}$U. The reason that thick seams of natural uranium ore are presently unable to undergo self-sustaining nuclear fission chain reactions, *i.e.*, $k_{eff} < 1.0$, is because the $^{235}$U/$^{238}$U ratio is too small. The $^{238}$U absorbs too high a proportion of neutrons. Because the half-life of $^{235}$U is shorter than that of $^{238}$U, the ratio of $^{235}$U/$^{238}$U was higher in the geological past, making possible the condition for natural fission, $k_{eff} \geq 1.0$.

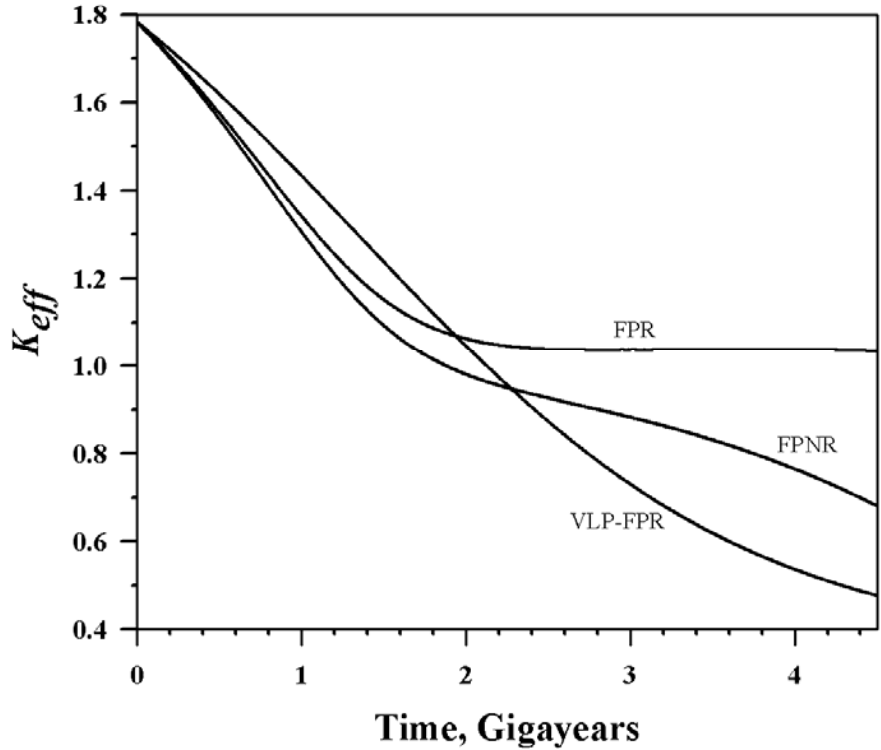

**Figure 3.** Numerical simulation results, chosen to illustrate main georeactor operational parameters and uncertainties, are presented in terms of $k_{eff}$ over the lifetime of the Earth. The curve labeled "VLP-FPR" is very low power for the case of fission products instantaneously removed. "FPR" is a 3 TW run also for the case of fission products instantaneously removed. "FPNR" is a 3 TW run with fission products not removed.

Main georeactor characteristic operational parameters and uncertainties are illustrated in Figure 3, showing $k_{eff}$ as a function of time for several numerical simulations made at constant fission powers. These show the importance of breeding, fission-product removal, and intrinsic self-regulation.

In Figure 3, the curve labeled "VLP-FPR" shows the necessity for breeding. In this example, the very-low-power fission produced only insignificant amounts of fissionable actinides. Consequently, the $k_{eff}$ was determined almost entirely by the natural decay of uranium, and, by the end of about 2 gigayears of operation, self-sustained nuclear fission chain reactions become impossible.

The "VLP-FPR" and the curve labeled "FPR" were calculated with instantaneous removal of fission products. But the "FPR" curve was calculated at a much higher power level where breeding kept $k_{eff} > 1.0$. As noted by Herndon (1994) and Seifritz (2003), the principal fuel-breeding takes place by the reaction $^{238}U(n,\gamma)^{239}U(\beta^-)^{239}Np(\beta^-)^{239}Pu(\alpha)^{235}U$. Too low an operating power will lead to insufficient breeding, whereas at power levels too high, the uranium fuel would be entirely consumed too early in the lifetime of the Earth.

For the georeactor to be able to operate into the present, fission products must be removed naturally. That necessity is shown quite clearly in Figure 3 by the curve labeled "FPNR", calculated with fission products not removed. After operation of about 1.5 gigayears, $k_{eff} < 1.0$ and self-sustained nuclear fission chain reactions become impossible. As I have discussed (Herndon, 1993, 1994), there is a natural mechanism for georeactor fission product removal: At the center of the Earth, density is a function almost entirely of atomic number and atomic mass. The fission process splits the actinide nucleus into two pieces, each being considerably less dense than its parent. At the high sub-core temperatures, even in the microgravity environment, these would tend to separate on the basis of density. This process may operate as one self-regulation mechanism.

Another, yet unknown, self-regulation mechanism appears evident from the curve labeled "FPNR" in Figure 3. Note that, at the time of Earth formation, the value of $k_{eff}$ is quite high; the uranium mix is "hot". In the numerical simulation, fission power generation was specified and controlled. In nature, without a self-regulation mechanism operating, at this high a value of $k_{eff}$, the georeactor would have run wild and might have burned out its uranium fuel long before life had existed on Earth. Early on, before about 1.5 gigayears of operation, fission product accumulation alone would not have been an effective self-control mechanism. Some other mechanism must have operated.

## 15. Radionuclide Abundance and Distribution

Much is yet unknown concerning the distribution of radionuclides within the Earth. Because of the identity between the parts of the endo-Earth and corresponding parts of the Abee enstatite chondrite, it is possible to make direct inferences as to radionuclide states of oxidation and locations within the endo-Earth, although not to the degree of precision that might ultimately be possible given adequate petrologic data with modern instrumentation and appropriate laboratory experiments (Herndon, 1998, 2005a). It is likewise possible to make some rough estimates of current georeactor energy production and uranium consumption, but past georeactor operation is for the most part unknown.

Within those limitations, the following generalizations concerning the endo-Earth radionuclides can be made: (*i*) Most of the $^{40}$K may be expected to exist in combination with oxygen in the silicates of the lower mantle, perhaps being confined to transition-region between the upper and the lower mantle; (*ii*) Uranium may be expected to exist at the center of the Earth where it may undergo self-sustaining nuclear fission chain reactions, but there is a possibility some non-fissioning uranium may be found scattered diffusely within the CaS core floaters; and, (*iii*) Thorium may be expected to occur within the core floaters at the core-mantle boundary, although its presence as well at the center of the Earth cannot be ruled out. Thorium is unable to be georeactor fuel or to be converted into fuel for the georeactor (Herndon and Edgerley, 2005).

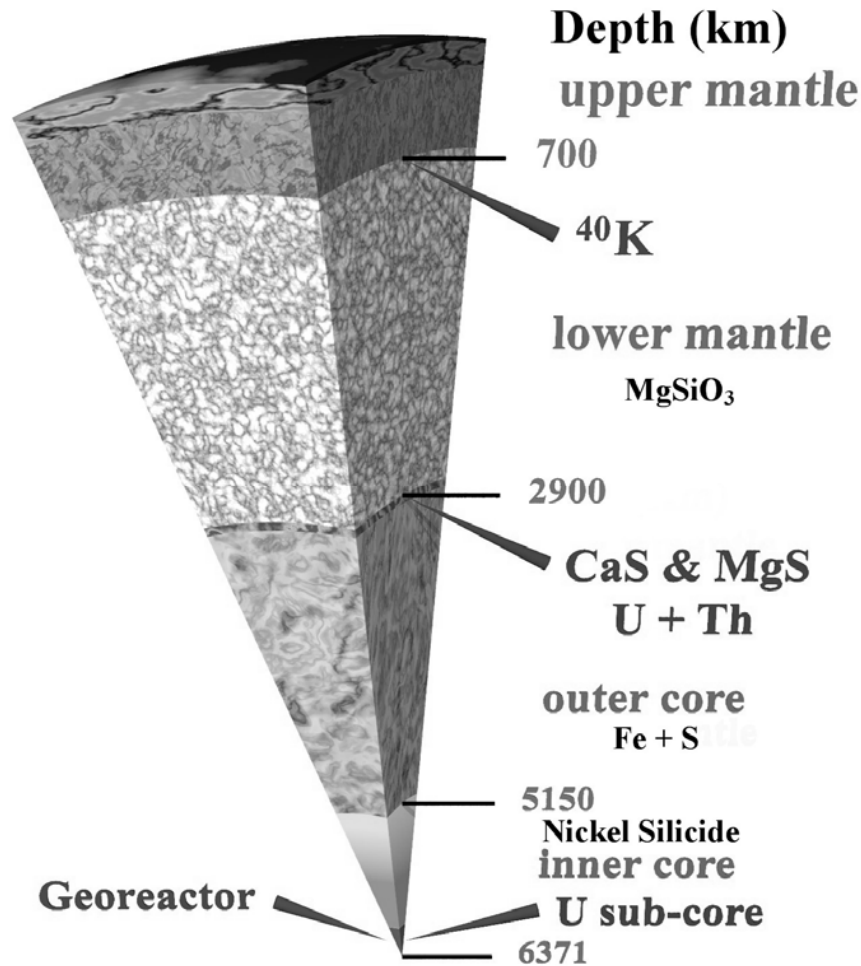

**Figure 4.** Schematic representation of the interior of the Earth showing regions in the endo-Earth where radionuclides may be expect to be concentrated.

Radionuclide abundance estimates for the endo-Earth and guesses for the exo-Earth are shown in Table 5. Their respective locations are represented schematically in Figure 4. In demonstrating the feasibility of the georeactor, I used very conservative uranium estimates, amounting to approximately 20% of the estimated total possible initial endo-Earth uranium content (Herndon, 1993; Hollenbach and Herndon, 2001). The results shown in Table 5, are based upon results of numerical simulations assuming that the entire amount of uranium is available for nuclear fission (Herndon and Edgerley, 2005). These, therefore, provide some boundary conditions on the maximum present-time radionuclide abundances.

**Table 5**. Estimates of the maximum present-day radionuclide content within the endo-Earth and guessed amounts in the exo-Earth. Endo-Earth values of uranium in parentheses, given for reference only, assume no fission activity. Data from (Baedecker and Wasson, 1975; Murrell and Burnett, 1982).

| Endo-Earth (estimate) | | |
|---|---|---|
| *Nuclide* | *Kilograms* | |
| $^{40}$K | $5.001 \times 10^{17}$ | |
| $^{232}$Th | $1.322 \times 10^{17}$ | |
| $^{235}$U | $3.065 \times 10^{14}$ | $(2.504 \times 10^{14})$ |
| $^{238}$U | $3.373 \times 10^{15}$ | $(3.456 \times 10^{16})$ |

| Exo-Earth (guess) | |
|---|---|
| *Nuclide* | *Kilograms* |
| $^{40}$K | $1.100 \times 10^{17}$ |
| $^{232}$Th | $2.908 \times 10^{16}$ |
| $^{235}$U | $4.629 \times 10^{15}$ |
| $^{238}$U | $1.528 \times 10^{16}$ |

In a series of numerical simulations run at successively higher power levels, Edgerley and I found that, with the same maximum initial endo-Earth uranium content, the georeactor could operate at a constant power level of as much as 30 TW and still be operating (Herndon and Edgerley, 2005). The question of power level, especially in times past, is the greatest unknown. Measurements of geo-antineutrinos pose the possibility of revealing the current distribution of radioactive nuclides and fission products.

## 16. Georeactor Variability

Seated deep within the Earth, the geomagnetic field varies in intensity and reverses polarity frequently, but quite irregularly, with an average time between reversals of about 200,000 years. Previously envisioned deep-Earth energy sources, including natural radioactivity, change only gradually and in only one direction over time. Variations, in the geomagnetic field, therefore, have previously only been ascribed to some mechanical instability in its production mechanism. I have suggested that the variable and intermittent changes in the intensity and direction of the geomagnetic field may have their origin in nuclear reactor variability (Herndon, 1993). Generally, variability in nuclear fission reactors arises from changes in composition and/or position of fuel, moderators, and neutron absorbers. Although as yet there is no irrefutable evidence of planetocentric nuclear reactor variability, circumstantial evidence certainly invites inquiry.

Upon considering observations of Jupiter's internally-generated energy, I demonstrated the feasibility of planetocentric nuclear fission reactors as energy sources for the giant planets (Herndon, 1992) in part using the same type of calculations employed by Kuroda (1956) to predict conditions for the natural reactors that were later discovered at Oklo, Republic of Gabon (Bodu et al., 1972; Fréjacques et al., 1975; Hagemann et al., 1975). The near-surface natural reactors at Oklo, which were critical about 1.8 gigayears ago, operated intermittently (Maurette, 1976). Recent investigations suggest quite rapid cycling periods with 0.5 hour of operation followed by 2.5 hours of dormancy (Meshik et al., 2004). While the specific control mechanism, presumably involving water, may not be directly applicable to the planetocentric reactors, the observations nevertheless demonstrate the potential variability of natural nuclear reactors.

Atmospheric turbulence in the giant planets appears to be driven by their internal energy sources. Jupiter, Saturn, and Neptune produce prodigious amounts of energy and display prominent turbulent atmospheric features. Uranus, on the other hand, radiates little, if any, internally generated energy and appears featureless. In the summer of 1878, Jupiter's Great Red Spot increased to a prominence never before recorded and, late in 1882, its prominence, darkness, and general visibility began declining so steadily that by 1890 astronomers thought that the Great Red Spot was doomed to extinction. Changes have been observed in other Jovian features, including the formation of a new lateral belt of atmospheric turbulence (Peek, 1958).

Jupiter, 98% of which consists of a mixture of H and He, an excellent heat transfer medium, is capable of rapid thermal transport. It is important to establish whether these atmospheric changes are due to changes in planetocentric nuclear reactor output as it seems, especially as these would represent short-period variability (Herndon, 1994). Ultimately, one may hope to understand the nature and possible variability of georeactor energy production by making fundamental discoveries and by discovering fundamental quantitative relationships in nature.

### 17. Deep-Earth Helium Evidence of the Georeactor

Clarke et al. (1969) discovered that $^3$He and $^4$He are venting from the Earth's interior. The $^3$He/$^4$He ratio of helium released to the oceans at the mid-oceanic ridges is about eight times greater than in the atmosphere (R/R$_A$ = 8 ±1, where R is the measured value of $^3$He/$^4$He and R$_A$ is the same ratio measured in air = 1.4 x 10$^{-6}$), and, therefore, cannot be ascribed to atmospheric contamination. High helium ratios, *e.g.*, ~37 R$_A$ (Hilton et al. 1999), have been observed from deep-source plumes, such as Iceland and Hawaii.

Previously, lacking knowledge of a deep-source $^3$He production mechanism, deep-Earth $^3$He has been assumed to be of primordial origin (Clarke et al., 1969; Hilton et al., 1999), trapped within the mantle at the time that the Earth formed. But the ratio of primordial $^3$He/$^4$He is thought to be ~$10^{-4}$, a value inferred from gas-rich meteorites (Pepin and Singer, 1965), which is about one order of magnitude greater than helium released from the mantle. In ascribing a primordial origin to the observed deep-Earth $^3$He/$^4$He, the assumption implicitly made is that the primordial component is diluted by a factor of about 10 with $^4$He produced by the natural radioactive decay of uranium and thorium in the mantle and/or in the crust. The alternative suggestion (Anderson, 1993), that the $^3$He/$^4$He arises instead from cosmic dust, subducted into the mantle, necessitates assuming that the influx of interplanetary dust particles was considerably greater in ancient times than at present and also assuming a ten-fold dilution by $^4$He.

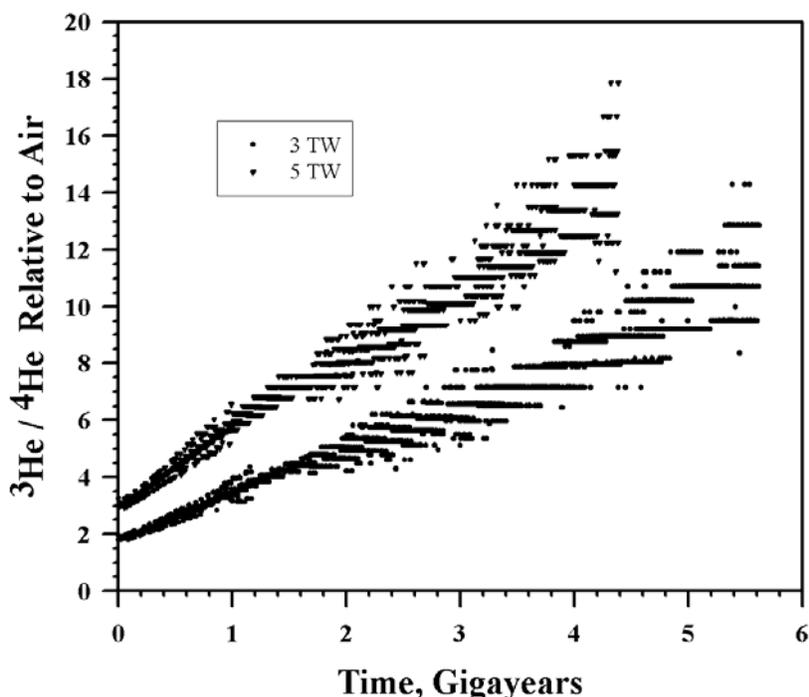

**Figure 5.** Nuclear georeactor numerical simulation results for 3 TW and 5 TW power levels showing the $^3$He/$^4$He ratios relative to air ($R_A$) produced during $2 \times 10^6$ year increments over the lifetime of the georeactor. Each data point represents the ratio of the $^3$He and $^4$He fission yields for a single time step. The pronounced upward trend of the data results from the continuing reduction of $^{238}$U, the principle source of $^4$He, by fission and by breeding. From (Herndon, 2003).

Helium isotope fission products from georeactor numerical simulations made at the Oak Ridge National Laboratory are shown in Figure 5. The data shown are values of the $^3$He/$^4$He ratio, relative to the same ratio in air ($R_A$), at each $2 \times 10^6$ year time step for each power level of the numerical simulations. For comparison, the range of values of the same ratio, measured in oceanic basalts, is shown in Table 6 at a 2σ confidence level. The entire range of $^3$He/$^4$He values from oceanic basalts, shown in Table 6, is produced by self-sustaining nuclear fission chain reactions, as demonstrated by the georeactor numerical simulations results presented in Figure 5.

I have suggested that the observed deep-source helium is georeactor-produced and is in fact strong evidence for the georeactor's existence (Herndon, 2003; Hollenbach and Herndon, 2001). Rao (2002) has provided background information and described the georeactor as being the solution to the riddles of relative abundances of helium isotopes and geomagnetic field variability.

**Table 6**. Statistics of $^3$He/$^4$He relative to air ($R_A$) of basalts from along the global spreading ridge system at a two standard deviation ($2\sigma$) confidence level. Adapted from Anderson (2000).

| | |
|---|---|
| Propagating Lithospheric Tears | $11.75 \pm 5.13\ R_A$ |
| Manus Basin | $10.67 \pm 3.36\ R_A$ |
| New Rifts | $10.01 \pm 4.67\ R_A$ |
| Continental Rifts or Narrow Oceans | $9.93 \pm 5.18\ R_A$ |
| South Atlantic Seamounts | $9.77 \pm 1.40\ R_A$ |
| MORB | $8.58 \pm 1.81\ R_A$ |
| EM Islands | $7.89 \pm 3.63\ R_A$ |
| North Chile Rise | $7.78 \pm 0.24\ R_A$ |
| Ridge Abandoned Islands | $7.10 \pm 2.44\ R_A$ |
| South Chile Rise | $6.88 \pm 1.72\ R_A$ |
| Central Atlantic Islands | $6.65 \pm 1.28\ R_A$ |
| HIMU Islands | $6.38 \pm 0.94\ R_A$ |
| Abandoned Ridges | $6.08 \pm 1.80\ R_A$ |

## 18. Eventual Demise of the Georeactor

Energy production by natural radioactive decay is predictable over time, decreasing gradually at known rates, and will continue to do so well into the future. By contrast, the consumption of uranium by georeactor-nuclear-fission may not have been constant in the past. At some point, the uranium supply of the georeactor may become exhausted, burned out by nuclear fission, possibly much sooner than it would have been exhausted by radioactive decay alone. The high $^3$He/$^4$He values in certain measurements of so-called plumes, specifically Icelandic and Hawaiian, may indicate the approach of the demise of the georeactor (Herndon, 2003).

In Figure 5, the upward trend over time of the data for each power level is principally the consequence of the diminishment by natural decay and by fuel breeding of $^{238}$U, the principle source of $^4$He. For a particular power level, the highest values represent the most recent production, especially near the end of the nuclear fission lifetime of the georeactor.

The limitation on the upper limits for $^3$He/$^4$He depends upon the georeactor being critical, *i.e.*, $k_{eff} \geq 1.0$, as its actinide fuel approaches depletion. The main factors affecting that circumstance are the amount and nature of the initial actinide sub-core and the operating history of the georeactor. One may reasonably expect, therefore, that the high values for $^3$He/$^4$He, shown in Figure 5, may not be true upper limits. It seems reasonable, though, that the high helium isotope ratios, measured in Hawaiian and in Icelandic basalt (Hilton et al. 1999), may signal the approach of the end of georeactor lifetime, although one may presently only speculate as to the time-frame involved.

One shortcoming of oceanic basalt helium isotopic measurements is that the time of formation of the helium is unknown. But from Figure 5, one can see that helium time-of-formation is important for assessing the time of demise of the georeactor. Future precision measurements of geo-antineutrinos may help to address that shortcoming.

### 19. Grand Overview and Generalizations

Only three processes, operant during the formation of the Solar System, are responsible for the diversity of compositions observed in planets, asteroids, and comets and are directly responsible for planetary internal-structures and dynamical processes, including and especially, geodynamics. These processes are: (*i*) Low-pressure, low-temperature condensation from solar matter in the remote reaches of the Solar System or in the interstellar medium; (*ii*) High-pressure, high-temperature condensation from solar matter associated with planetary-formation by raining out from the interiors of giant-gaseous protoplanets, and; (*iii*) Stripping of the primordial volatile components from the inner portion of the Solar System by super-intense solar wind associated with T-Tauri phase mass-ejections, presumably during the thermonuclear ignition of the Sun.

Low-pressure, low-temperature condensation from solar matter in the remote reaches of the Solar System or in the interstellar medium is the process responsible for cometary matter, and is responsible for one of the two components from which ordinary chondrite meteorites are composed. It is responsible for the primitive Orgueil-like carbonaceous chondrite meteorites and, after separation from primordial volatile components and being melted and/or re-evaporated and re-condensed, it is responsible for the more crystallized and evolved carbonaceous chondrites, such as the Allende meteorite. This type of matter contributes to the terrestrial planets only as a late-addition veneer component.

High-pressure, high-temperature condensation from solar matter, associated with planetary-formation by raining out from the interiors of giant-gaseous protoplanets, is the process responsible for the bulk of planetary formation and for establishing the highly reduced state of oxidation of planetary interiors. Internal planetary structures are produced as a consequence of the highly reduced state of planetary interiors, including the occurrence of major quantities of uranium and thorium in planetary cores, leading to planetocentric nuclear fission reactors. That same condensation process is responsible for Earth formation as a giant gaseous Jupiter-like planet and for storing vast amounts of the energy of protoplanetary compression in the rock-plus-alloy kernel that became Earth as we know it.

Stripping of the primordial volatile components from the inner portion of the Solar System by super-intense solar wind associated with T-Tauri phase mass-ejections, presumably during the thermonuclear ignition of the Sun, is the process responsible for removal of any gaseous components that might have been associated with the formation of terrestrial planets, including removal of part of the protoplanet of Mercury, which became the other of the two components from which ordinary-chondrite matter formed in the region of the asteroid belt. It is the process responsible removing approximately 300 Earth-masses of primordial volatile gases from the Earth, which began Earth's decompression process, making available vast amounts of energy for driving geodynamic processes which I have described by the new whole-Earth decompression dynamics, and which is responsible for emplacing heat at the mantle-crust-interface at the base of the crust through the process I have described, called mantle decompression thermal-tsunami.

The three processes, operant during the formation of the Solar System, lead logically, in a causally related manner, to a coherent vision of planetary formation with profound implications. Consequently, there is reason to suppose that each planet and, perhaps, each of the larger moons, has at its center, a region of highly reduced enstatite-chondrite-like matter and a uranium sub-core at one time capable of self-sustained nuclear fission reactions. The vision of planetary formation presented here is consistent with observations of near-to-star gas-giants in other planetary systems. The geodynamic processes for the terrestrial planets may differ from one another to some extent, not so much due to their interiors, but as a consequence of the circumstances of their accumulation and removal of primordial volatile components.

These are exciting times in the natural physical sciences. Along with the new understanding of Solar System formation and whole-Earth geodynamics described above, new experimental advances are being made that, I submit, will inevitably confirm and perhaps extend these concepts. Already, astronomers are beginning to image remote planetary systems and finding close-to star gas giants like Earth at a very early stage. Neutrino physicists, with decades of experience measuring neutrinos from the Sun and from outer space, are beginning to detect anti-neutrinos from within our own planet. To image the interior of the Earth using anti-neutrinos, physicists face great challenges in attempting to attain the high resolution and directionality needed. But facing great challenges and making important discoveries is what science is all about.